\documentclass[runningheads]{llncs}
\usepackage{graphicx}
\usepackage{amsmath}
\usepackage{hyperref}
\usepackage{longtable}
\usepackage{multirow}
\usepackage{chngpage}% allows for temporary adjustment of side margins
\usepackage{times}

\makeatletter
\newcommand{\printfnsymbol}[1]{%
  \textsuperscript{\@fnsymbol{#1}}%
}
\makeatother

\begin{document}
\title{A multispeaker dataset of raw and reconstructed speech production real-time MRI video and 3D volumetric images\thanks{Article submitted to Nature Scientific Data.}}

\titlerunning{ }

\author{Yongwan Lim\inst{1}\thanks{YL and AT should be considered joint first-authors} \and
Asterios Toutios\inst{1}\printfnsymbol{2}\and
Yannick Bliesener\inst{1} \and
Ye Tian\inst{1} \and
Sajan Goud Lingala\inst{1} \and
Colin Vaz\inst{1} \and
Tanner Sorensen\inst{2} \and
Miran Oh\inst{2} \and
Sarah Harper\inst{2} \and
Weiyi Chen\inst{1} \and
Yoonjeong Lee\inst{2} \and
Johannes Töger\inst{1} \and
Mairym Lloréns Montesserin\inst{2} \and
Caitlin Smith\inst{2} \and
Bianca Godinez\inst{3} \and
Louis Goldstein\inst{2} \and
Dani Byrd\inst{2} \and
Krishna S. Nayak\inst{1}\and
Shrikanth Narayanan\inst{1,2}\thanks{Corresponding author(s): Shrikanth S. Narayanan (shri@sipi.usc.edu)}}

\date{\today}
% \author{Yongwan Lim\inst{1}\orcidID{0000-1111-2222-3333} \and
% Asterios Toutios\inst{1}\orcidID{} \and
% Yannick Bliesener\inst{1}\orcidID{} \and
% Ye Tian\inst{1}\orcidID{} \and
% Sajan Goud Lingala\inst{1}\orcidID{} \and
% Colin Vaz\inst{1}\orcidID{} \and
% Tanner Sorensen\inst{2}\orcidID{} \and
% Miran Oh\inst{2}\orcidID{} \and
% Sarah Harper\inst{2}\orcidID{} \and
% Weiyi Chen\inst{1}\orcidID{} \and
% Yoonjeong Lee\inst{2}\orcidID{} \and
% Johannes Töger\inst{1}\orcidID{} \and
% Mairym Lloréns Montesserin\inst{2}\orcidID{} \and
% Caitlin Smith\inst{2}\orcidID{} \and
% Bianca Godinez\inst{3}\orcidID{} \and
% Louis Goldstein\inst{2}\orcidID{} \and
% Dani Byrd\inst{2}\orcidID{} \and
% Krishna S. Nayak\inst{1}\orcidID{2222--3333-4444-5555}\and
% Shrikanth Narayanan\inst{1,2}\orcidID{1111-2222-3333-4444}}
% \author{Yongwan Lim\inst{1}\orcidID{0000-1111-2222-3333} \and
% Shrikanth Narayanan\inst{2,3}\orcidID{1111-2222-3333-4444} \and
% Krishna S. Nayak\inst{3}\orcidID{2222--3333-4444-5555}}
%
\authorrunning{ }

\institute{Ming Hsieh Department of Electrical and Computer Engineering, \\ Viterbi School of Engineering, University of Southern California, Los Angeles, CA, USA \and
Department of Linguistics, Dornsife College of Letters, Arts and Sciences, University of Southern California, Los Angeles, CA, USA \and
Department of Linguistics, California State University Long Beach, Long Beach, CA, USA}

\maketitle              % typeset the header of the contribution
\begin{abstract}
Real-time magnetic resonance imaging (RT-MRI) of human speech production is enabling significant advances in speech science, linguistics, bio-inspired speech technology development, and clinical applications. Easy access to RT-MRI is however limited, and comprehensive datasets with broad access are needed to catalyze research across numerous domains. The imaging of the rapidly moving articulators and dynamic airway shaping during speech demands high spatio-temporal resolution and robust reconstruction methods. Further, while reconstructed images have been published, to-date there is no open dataset providing raw multi-coil RT-MRI data from an optimized speech production experimental setup. Such datasets could enable new and improved methods for dynamic image reconstruction, artifact correction, feature extraction, and direct extraction of linguistically-relevant biomarkers. The present dataset offers a unique corpus of 2D sagittal-view RT-MRI videos along with synchronized audio for 75 subjects performing linguistically motivated speech tasks, alongside the corresponding first-ever public domain raw RT-MRI data. The dataset also includes 3D volumetric vocal tract MRI during sustained speech sounds and high-resolution static anatomical T2-weighted upper airway MRI for each subject.

% \keywords{First keyword  \and Second keyword \and Another keyword.}
\end{abstract}

\section*{Background \& Summary}
Human upper airway functions such as swallowing, breathing, and speech production are the result of a well-coordinated choreography of various mobile soft tissue and muscular structures such as the tongue, lips, and velum, as well as bony structures such as the plate, mandible, and hyoid \cite{1,2}. The complexity and sophistication of human speech production poses a multitude of open research questions with implications for linguistics and speech science, as well as clinical and technological applications, creating a demand for improved methods for observing and measuring the vocal instrument in action \cite{3,4}. 

Real-Time Magnetic Resonance Imaging (RT-MRI) with concurrent audio recording has emerged as an imaging modality that can provide new insights into speech production with all its inherent systematicities and variation between languages, contexts, and individuals \cite{1,3,5,6}. This technique has the unique advantage of monitoring the complete vocal tract safely and non-invasively at relatively high spatial and temporal resolution. Applications of RT-MRI span multiple realms of research including study of: (i) phonetic and phonological phenomena, (ii) spoken language acquisition and breakdown, including the assessment and remediation of speech disorders \cite{1,3}; (iii) dynamics of vocal tract shaping during communicative speech or vocal performance; (iv) articulatory modeling and motor control; (v) speech synthesis and recognition technologies, and (vi) speaker modeling and biometrics. 

Several examples of recently published MRI data repositories \cite{7,8,9,10,11} have demonstrated the value of publicly available datasets to address a multitude of open challenges not only in scholarly research but also in translation into clinical and scientific applications. Open raw datasets have been used to validate and refine advanced algorithms and to train, validate, and benchmark promising recent applications of ideas inspired by artificial intelligence/machine learning methods \cite{12}. They are also valued by commercial vendors to showcase performance and generalizability of reconstruction methods. Despite a recognized impact in, for example, the domains of musculoskeletal and brain MRI, there are currently no raw MRI datasets involving the moving vocal tract. 

The vocal tract contains multiple rapidly moving articulators, which can change position significantly on a millisecond timescale – an imaging challenge necessitating high temporal resolution adequate for observing these dynamic speech process \cite{6}. While under-sampling of MRI measurements on time-efficient trajectories enables the desired resolution, such measurements are hampered by prolonged computation time for advanced image reconstruction, low signal-to-noise-ratio, and artifacts due to under-sampling and/or rapid differences of magnetic susceptibility \cite{13,14,15} at the articulator boundaries, which are of utmost interest in characterizing speech production. These limitations often render present day RT-MRI’s operating point beneath application demands and can introduce bias and increased variance during data analysis. Thus, there is much room for improvements in the imaging technology pipeline. Despite the promise of deep learning/machine learning approaches to provide superior performance both in reconstruction time and image quality, their application to dynamic imaging of fast aperiodic motion with high spatiotemporal resolution and low latency is still in its infancy. We speculate that this is in large measure due to the lack of large-scale public MRI datasets – the cornerstone of machine learning. 

This paper presents a unique dataset that offers videos of the entire vocal tract in action, with synchronized audio, imaged along the sagittal plane from 75 subjects while they performed a variety of speech tasks. The dataset also includes 3D volumetric vocal tract MRI during sustained speech sounds and high resolution static anatomical T2-weighted upper airway MRI for each subject. Unlike other open datasets for dynamic speech RT-MRI \cite{16,17,18,19,20}, the present dataset includes raw, multi-receiver-coil MRI data with non-Cartesian, spiral sampling trajectory. 

The dataset can be used to aid the development of algorithms that monitor fast aperiodic dynamics of speech articulators at high spatiotemporal resolution while offering simultaneous suppression of noise and artifacts due to sub-Nyquist sampling or susceptibility. The inclusion of an unprecedented number of 75 speakers can also help improve our scientific understanding of how vocal tract morphology and speech articulation interact and shed light on the stable and variable aspects of speech signal properties across speakers, providing for improved models of speech production in linguistics and speech science research. 

In sum, it is our hope and anticipation that the public and free provision of this rich dataset can further foster and stimulate research and innovation in the science of human speech production and its imaging.

\section*{Methods}
\subsection*{Participants}
Seventy-five healthy subjects (40 females and 35 males; 49 native and 26 non-native American English speakers; age 18-59 years) were included in this study. Each participant filled out a questionnaire on basic demographic information including birthplace, cities raised and lived, and first and second languages. Demographics are summarized in Table \ref{tab1}. All subjects had normal speech, hearing, and reading abilities, and reported no known physical or neurological abnormalities. All participants were cognizant of the nature of the study, provided written informed consent, and were scanned under a protocol approved by the Institutional Review Board of the University of Southern California (USC). The data were collected at the Los Angeles County – USC Medical Center between January 24, 2016 and February 24, 2019. \\ \\ \\ 

\clearpage
\renewcommand{\arraystretch}{1.05}
\begin{longtable}{cccccc}
    % \begin{adjustwidth}{-0.1in}{-0.1in} 
    \caption{Demographic information of subjects; Age at the time of scan (years), F: Female, M: Male, L1: first language, L2: second language (if any)}\label{tab1}  \\
    \hline
    \textbf{Subject} & \multirow{2}{*}{\textbf{Sex}} & \multirow{2}{*}{\textbf{Age}}	& \multirow{2}{*}{\textbf{L1}} & \multirow{2}{*}{\textbf{L2}} & \multirow{2}{*}{\textbf{City raised}} \\
    \textbf{ID} &&&&& \\
    \hline
    \endfirsthead
    
    \multicolumn{6}{c}%
    {\tablename\ \thetable\ $.$ \textit{Continued}} \\
    \hline
    \textbf{Subject} & \multirow{2}{*}{\textbf{Sex}} & \multirow{2}{*}{\textbf{Age}}	& \multirow{2}{*}{\textbf{L1}} & \multirow{2}{*}{\textbf{L2}} & \multirow{2}{*}{\textbf{City raised}} \\
    \textbf{ID} &&&&& \\
    \hline
    \endhead
    \hline 
    % \multicolumn{6}{c}{\textit{Continued on next page}} \\
    \endfoot
    \hline
    \endlastfoot
    
    sub001	& F &	24 &	Mandarin &	English &	Hohhot, China \\
    sub002	&F	&26&	English	&-&	Phoenix, AZ\\
    sub003	&F	&19&	English	&Mandarin&	Alhambra, CA\\
    sub004	&M	&21&	English	&-	&Portland, OR\\
    sub005	&F&	27	&English&	Spanish&	Ashland, KY\\
    sub006	&M&	27&	English&	-	&Exton, PA\\
    sub007	&F&	30&	English&	Spanish&	Alameda, CA\\
    sub008	&M&	32&	Telugu&	Hindi	&Hyderabad, India\\
    sub009	&M&	27&	English&	-	&Unknown\\
    sub010	&F&	21&	English	&-	&Cranston, RI\\
    sub011	&F&	19&	English	&German&	Ridgefield, CT\\
    sub012	&F&	27&	German	&English&	Siegen, Germany\\
    sub013	&M&	23&	English	&-	&Monrovia, CA\\
    sub014	&F&	26&	English	&Spanish&	Riverside, CA\\
    sub015	&M&	26&	English	&-	&Edison, NJ\\
    sub016	&F&	20&	English	&French&	Gurnee, IL\\
    sub017	&F&	31&	English	&-	&Queens, NY\\
    sub018	&F&	21&	English	&Italian&	West Newbury, MA\\
    sub019	&M&	26&	English	&-	&Virginia Beach, VA\\
    sub020	&M&	Unknown&	English&	Spanish&	Los Angeles, CA\\
    sub021	&F&	29&	Korean	&English&	Karachi, Pakistan\\
    sub022	&M&	49&	Tamil	&English&	New Delhi, India\\
    sub023	&M&	18&	English	&Spanish&	San Clemente, CA\\
    sub024	&F&	22&	Vietnamese&	English&	Austin, TX\\
    sub025	&M&	21&	English	&Mandarin&	San Jose, CA\\
    sub026	&F&	21&	English	&Bahasa Indonesia&	Jakarta, Indonesia\\
    sub027	&F&	21&	English	&Native Hawaiian&	Kaneohe, HI\\
    sub028	&M&	25&	English	&Hindi	&Mumbai, India\\
    sub029	&M&	42&	Greek	&English&	Serres, Greece\\
    sub030	&M&	28&	Gujarati&	English&	Rajkot, India\\
    sub031	&F&	20&	English	&-	&Thousand Oaks, CA\\
    sub032	&F&	21&	Korean	&English&	Seoul, S. Korea\\
    sub033	&M&	18&	English	&French	&San Diego, CA\\
    sub034	&M&	28&	Telugu/Kannada&	English&	Bangalore, India\\
    sub035	&M&	21&	English	&Spanish	&Castro Valley, CA\\
    sub036	&F&	25&	English	&German	&Norfolk, NE\\
    sub037	&M&	32&	Korean	&English&	Seoul, S. Korea\\
    sub038	&M&	30&	Russian	&English&	Novosibirsk, Russia\\
    sub039	&M&	27&	Chinese	&English&	Shanghai, China\\
    sub040	&M&	26&	Korean	&English&	Seoul, S. Korea\\
    sub041	&F&	22&	English	&French	&Nanjing, China\\
    sub042	&F&	22&	English&	Cantonese&	Monterey Park, CA\\
    sub043	&F&	28&	English	&Russian&	Houston, TX\\
    sub044	&F&	24&	English	&Spanish	&Atlanta, GA\\
    sub045	&M&	23&	English	&-	&Los Angeles, CA\\
    sub046	&M&	49&	English	&Spanish&	Los Angeles, CA\\
    sub047	&F&	18&	English	&-	&Sacramento, CA\\
    sub048	&F&	27&	English	&Spanish&	Manhattan Beach, CA\\
    sub049	&F&	29&	English	&-	&Sacramento, CA\\
    sub050	&M&	30&	English	&Korean&	Barrington, RI\\
    sub051	&M&	33&	Korean	&English	&Seoul, S. Korea\\
    sub052	&M&	26&	English	&German	&Omaha, NE\\
    sub053	&M&	29&	English	&-	&Mill Valley, CA\\
    sub054	&F&	24&	English	&Spanish&	Oswego, IL\\
    sub055	&F&	21&	English	&French	&Cincinnati, OH\\
    sub056	&M&	18&	Portuguese&	English&	Diamond Bar, CA\\
    sub057	&M&	19&	English	&Spanish	&Abington, PA\\
    sub058	&F&	28&	Spanish/English&	Italian&	San Juan, Pueto Rico\\
    sub059	&F&	22&	English	&Spanish	&Chantilly, VA\\
    sub060	&F&	22&	Korean	&English	&San Diego, CA\\
    sub061	&F&	26&	Tamil	&Hindi	&Ooty, India\\
    sub062	&M&	25&	English	&Twi	&Tema, Ghana\\
    sub063	&F&	26&	English	&Mandarin&	Charlestown, IN\\
    sub064	&F&	Unknown&	Telugu&	English&	Vijayawada, India\\
    sub065	&M&	19&	Spanish/English&	-&	New York, NY\\
    sub066	&F&	18&	Spanish/English&	-&	Los Angeles, CA\\
    sub067	&M&	59&	English	&French	&Washington, DC\\
    sub068	&M&	28&	Gujarati&	Hindi&	Junagadh, India\\
    sub069	&F&	25&	Mandarin&	English&	Suzhou, China\\
    sub070	&F&	29&	English	&-	&Brawley, CA\\
    sub071	&F&	25&	English	&Spanish&	Fresno, CA\\
    sub072	&M&	36&	English	&-	&Medina, OH\\
    sub073	&M&	30&	Korean	&English&	Iksan, S. Korea\\
    sub074	&F&	30&	Telugu	&Hindi	&Hyderabad, India\\
    sub075	&F&	29&	Spanish	&English&	Baldwin Park, CA\\
    % \end{tabular}
    % \end{adjustwidth}
\end{longtable}

\clearpage
\subsection*{Experimental Overview}
\begin{figure}[!hp]
    \begin{adjustwidth}{-0.1in}{-0.1in} 
	\centering
	\includegraphics[width=1\textwidth]{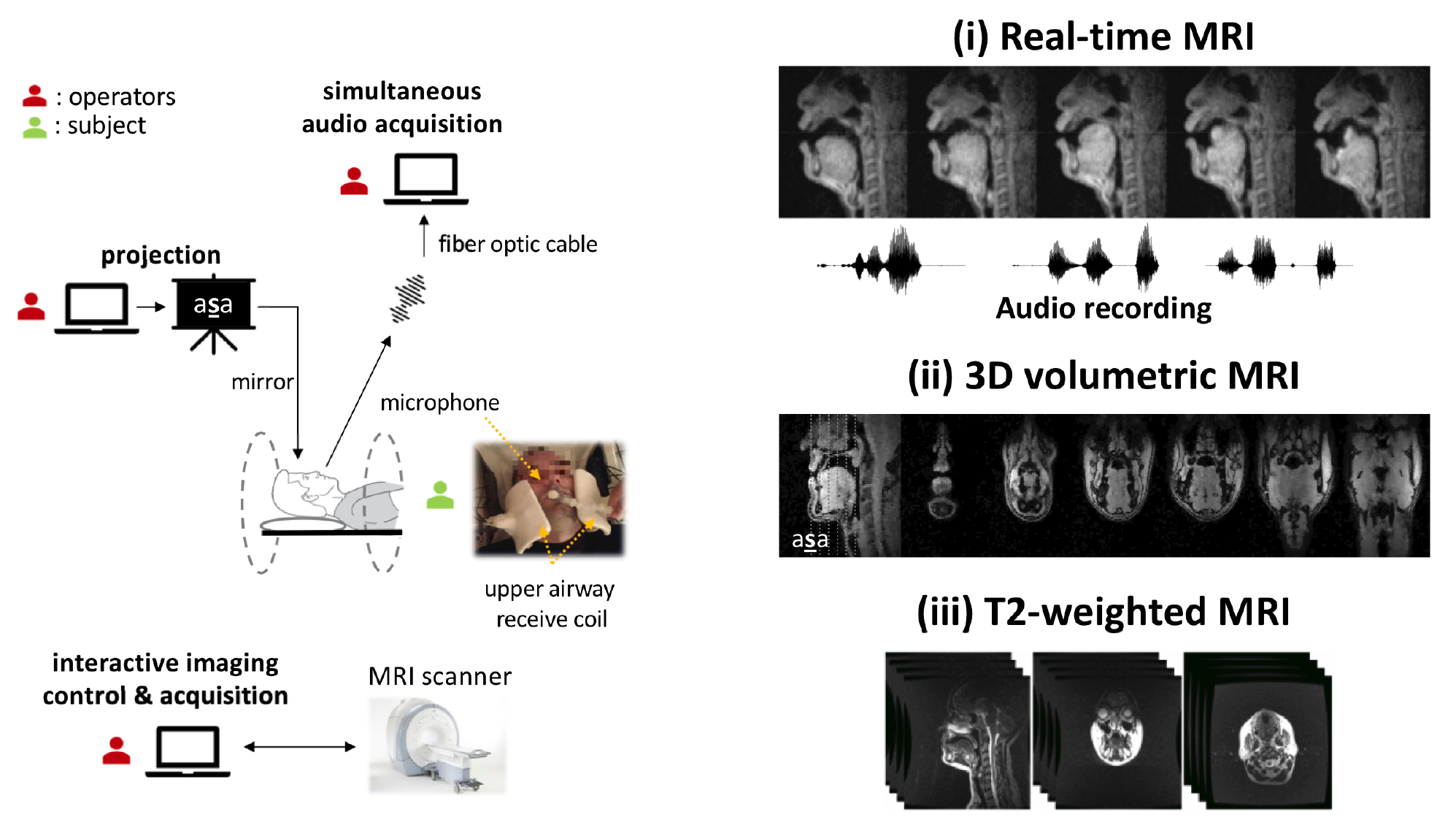}
	\caption{Data acquisition workflow and data records. (Left) Data were acquired at 1.5 Tesla using the custom upper-airway coil located in close proximity to the subject’s upper airway structures. The subject visualized the stimuli through a mirror-projector setup and audio was recorded through an MR-compatible microphone simultaneous with the RT-MRI. The scanner operator used a custom interactive imaging interface with the scanner hardware to control and acquire the data for the RT-MRI session. (Right) The recorded MRI data were: (i) dynamic, 2D real-time MRI of the vocal tract’s mid-sagittal slice at 83 frames per second during production of a comprehensive set of scripted and spontaneous speech material, along with synchronized audio recording; (ii) static, 3D volumetric images of the vocal tract, captured during sustained production of sounds or postures, 7 seconds each; (iii) T2-weighted volumetric images at rest position, capturing fine detail anatomical characteristics of the vocal tract.}
	\label{fig1}
	\end{adjustwidth}
\end{figure}

Three types of MRI data were recorded for each subject: (i) dynamic, real-time MRI of the vocal tract’s mid-sagittal slice at 83 frames per second during production of a comprehensive set of scripted and spontaneous speech material, averaging 17 minutes per subject, along with synchronized audio; (ii) static, 3D volumetric images of the vocal tract, captured during sustained production of sounds from the full set of American English vowels and continuant consonants, 7 seconds each; (iii) T2-weighted volumetric images at rest position, capturing fine detail anatomical characteristics of the vocal tract (See Figure \ref{fig1}). 

All data were collected using a commercial 1.5 Tesla MRI scanner (Signa Excite, GE Healthcare, Waukesha, WI) with gradients capable of 40 mT/m amplitude and 150 mT/m/ms slew rate. A custom 8‐channel upper airway receiver coil array \cite{21}, with four elements on each side of the subject’s cheeks, was used for signal reception. Compared to commercially available coils that are designed for neurovascular or carotid artery imaging, this custom coil has been shown to provide 2-fold to 6-fold higher signal-to-noise-ratio (SNR) efficiency in upper airway vocal tract regions of interests including tongue, lips, velum, epiglottis, and glottis. The subjects, while imaged, were presented with scripts of the experimental stimuli via a projector-mirror setup \cite{22}. Acoustic audio data were recorded inside the scanner using commercial fiber-optic microphones (Optoacoustics Inc., Yehuda, Israel) concurrently with the RT-MRI data acquisition using a custom recording setup \cite{23}.

\subsection*{Stimuli and Linguistic Justification}
The stimuli were designed to efficiently capture salient, static and dynamic, and articulatory and morphological aspects of speech production of American English in a single 90-minute scan session. 

Table \ref{tab2} lists the speech stimuli used for the RT-MRI data collection of the first 45-minute sub scan session. Each individual task was designed to be performed within 30 seconds at a normal speaking rate, however the actual recordings varied in duration depending on the length of the task and the natural speaking rate of the individual. The stimulus set contained material to elicit both scripted speech and spontaneous speech. The scripted speech tasks were consonant production in symmetric vowel-consonant-vowel context, vowels /V/ produced between the consonants /b/ and /t/, i.e., in /bVt/ contexts, four phonologically rich sentences \cite{24}, and three reading passages commonly used in speech evaluation and linguistic studies \cite{25,26,27}. Scripted instructions to produce several gestures were also included. These gestures were clenching, wide opening of mouth, yawning, swallowing, slow production of the sequence /i/-/a/-/u/-/i/, tracing of the palate with the tongue tip, and singing “la” at their highest and lowest pitches. The stimuli in the scripted speech were repeated twice. The spontaneous speech tasks were to describe the content and context of five photographs (Supplemental Figure \ref{supfig1}) and to answer five open-ended questions (e.g., “What is your favorite music?”). The full scripts and questions are provided in Table \ref{tab3}.

% \setlength\LTleft{0pt}
% \setlength\LTright{0pt}
% \begin{longtable}{ccccc}
%     % \begin{adjustwidth}{-0.37in}{-0.37in} 
%     % \captionsetup{justification=centering}
%     \caption{Speech experiment stimuli for 2D RT-MRI.}    \label{tab2}
%     % \centering
%     \hline
%     \multirow{2}{*}{\textbf{Category}} &	\textbf{Stimulus}	&\textbf{Stimulus}  &	\multirow{2}{*}{\textbf{Description}} &	\textbf{Duration}  \\ 
%      &	\textbf{Index}	& \textbf{Name} &	 &	 (\textbf{sec}) \\ \hline
%     \endhead
%     & 01, 02, 03& 	vcv[1-3]& 	Consonants in symmetric /VCV/	& 30 (x3)\\
% 	& 04& 	bvt& 	Vowels in /bVt/	& 30\\
% 	 & 05& 	shibboleth	& Four phonologically rich sentences \cite{34} & 30\\
% 	Scripted& 06& 	rainbow& 	Rainbow passage \cite{35} & 30\\
% 	speech& 07, 08	& grandfather[1-2]	& Grandfather passage \cite{36} & 30 (x2)\\
% 	& 09, 10	& northwind[1-2]	& Northwind and the sun passage \cite{37} & 30 (x2)\\
% 	& 11	& postures	& Postures	& 30\\
% 	\cline{2-5}
% 	& \multicolumn{3}{c}{Repetition of the above scripted speech tasks}	& 30 (x11)\\ \hline
%     Spontaneous 	&12 – 16&	picture[1-5]&	Description of pictures	& 30 (x5) \\
% 	speech & 17 – 21	&topic[1-5]	&Discussion about topics &	30 (x5) \\ \hline

%     % \end{adjustwidth}
% \end{longtable}
\clearpage
\renewcommand{\arraystretch}{1.3}
\begin{table}[!h]
    \begin{adjustwidth}{-0.37in}{-0.37in} 
    % \captionsetup{justification=centering}
    \caption{Speech experiment stimuli for 2D RT-MRI.}
    \centering
    \begin{tabular}{ccccc}
    \hline
    \multirow{2}{*}{\textbf{Category}} &	\textbf{Stimulus}	&\textbf{Stimulus}  &	\multirow{2}{*}{\textbf{Description}} &	\textbf{Duration}  \\ 
     &	\textbf{Index}	& \textbf{Name} &	 &	 (\textbf{sec}) \\ \hline
    & 01, 02, 03& 	vcv[1-3]& 	Consonants in symmetric /VCV/	& 30 (x3)\\
	& 04& 	bvt& 	Vowels in /bVt/	& 30\\
	 & 05& 	shibboleth	& Four phonologically rich sentences \cite{34} & 30\\
	Scripted& 06& 	rainbow& 	Rainbow passage \cite{35} & 30\\
	speech& 07, 08	& grandfather[1-2]	& Grandfather passage \cite{36} & 30 (x2)\\
	& 09, 10	& northwind[1-2]	& Northwind and the sun passage \cite{37} & 30 (x2)\\
	& 11	& postures	& Postures	& 30\\
	\cline{2-5}
	& \multicolumn{3}{c}{Repetition of the above scripted speech tasks}	& 30 (x11)\\ \hline
    Spontaneous 	&12 – 16&	picture[1-5]&	Description of pictures	& 30 (x5) \\
	speech & 17 – 21	&topic[1-5]	&Discussion about topics &	30 (x5) \\ \hline

    \end{tabular}
    \label{tab2}
    \end{adjustwidth}
\end{table}

\renewcommand{\arraystretch}{1.05}
\begin{table}[!p]
    \begin{adjustwidth}{-0.37in}{-0.37in}  
    \caption{Full scripts and questions of speech experiment stimuli for 2D RT-MRI.}
    \centering
    \begin{tabular}{cl}
    \hline
    \textbf{Stimulus}	&	\multicolumn{1}{c}{\multirow{2}{*}{\textbf{Script or question/topic}}} \\ 
    \textbf{Index}	&  \\ \hline

    vcv1 &	apa upu ipi ata utu iti aka uku iki aba ubu ibi ada udu idi aga ugu igi \\\hline
    vcv2&	atha uthu ithi asa usu isi asha ushu ishi ama umu imi ana unu ini ala ulu ili \\\hline
    vcv3&	afa ufu ifi ava uvu ivi ara uru iri aha uhu ihi awa uwu iwi aya uyu iyi \\\hline
    bvt	& beet bit bait bet bat pot but bought boat boot put bite beaut bird boyd abbot\\ \hline
	\multirow{5}{*}{shibboleth}	& She had your dark suit in greasy wash water all year.\\
    & Don't ask me to carry an oily rag like that.\\
    & The girl was thirsty and drank some juice followed by a coke.\\
    & Your good pants look great however your ripped pants look like a cheap version\\
    & of a k-mart special is that an oil stain on them. \\\hline
    \multirow{7}{*}{rainbow}	
    & When the sunlight strikes raindrops on the air they act as a prism and form \\
    & a rainbow. The rainbow is a division of white light into many beautiful colors.\\
    & These take the shape of a long round arc with its path high above and its two \\
    & ends apparently beyond the horizon.  There is according to legend a boiling pot\\
    & of gold at one end. People look but no one ever finds it. When a man looks for \\
    & something beyond his reach his friends say he is looking for the pot of gold \\
    & at the end of the rainbow. \\\hline
    
    \multirow{5}{*}{grandfather1} 
    & You wish to know all about my grandfather. Well, he is nearly ninety three \\
    & years old yet he still thinks as swiftly as ever. He dresses himself in an \\
    & old black frock coat usually several buttons missing.  A long beard clings \\
    & to his chin, giving those who observe him a pronounced feeling of the utmost \\
    & respect. When he speaks, his voice is just a bit cracked and quivers a bit. \\ \hline
    \multirow{5}{*}{grandfather2}
    & Twice each day he plays skillfully and with zest upon a small organ. \\
    & Except in the winter when the snow or ice prevents, he slowly takes a short \\
    & walk in the open air each day.  We have often urged him to walk more and \\
    & smoke less but he always answers, “banana oil” grandfather likes to be modern \\
    & in his language. \\ \hline
    \multirow{4}{*}{northwind1}
    & The north wind and sun were disputing which was the stronger, when a traveler \\
    & came along wrapped in a warm cloak. They agreed that the one who first \\
    & succeeded in making the traveler take off his cloak should be considered \\
    & stronger than the other. \\ \hline
    \multirow{5}{*}{northwind2}
    & Then the north wind blew as hard as he could, but the more he blew \\
    & the more closely did the traveler fold his cloak around him and at last \\
    & the north wind gave up the attempt.  Then the sun shone out warmly and \\ 
    & immediately the traveler took off his cloak and so the north wind was \\
    & obliged to confess that the sun was stronger of the two. \\ \hline
    \multirow{3}{*}{postures}
    & clench, open wide \& yawn, swallow, eee\dots aahh\dots uuww\dots eee, \\
    & trace palate with tongue tip, Sing “la” at your highest note, \\
    & Sing “la” at your lowest note \\ \hline
    
    topic1	&My favorite music \\
    topic2	&How do I like LA?\\
    topic3	&My favorite movie\\
    topic4	&Best place I've been to\\
    topic5	&My favorite restaurant\\ \hline

    \end{tabular}
    \label{tab3}
    \end{adjustwidth}
\end{table}

Table \ref{tab4} lists the speech stimuli used for the 3D volumetric data collection of a 30-minute part of the scan session. Each individual task was designed for a subject to sustain vowels, consonant sounds, or to maintain postures for 7 seconds. The stimulus set contained consonant production in symmetric vowel-consonant-vowel context, vowels /V/ produced between the consonants /b/ (or /p/) and /t/, and production of several postures. All of the stimuli were repeated twice.

\renewcommand{\arraystretch}{1.05}
\begin{table}
    \begin{adjustwidth}{-0.37in}{-0.37in} 
    \caption{Speech experiment stimuli for 3D volumetric static MRI.}
    \centering
    \begin{tabular}{ccccc}
    \hline
    \multirow{2}{*}{\textbf{Category}} &	\textbf{Stimulus}	&\textbf{Stimulus}  &	\multirow{2}{*}{\textbf{Description}} &	\textbf{Duration}  \\ 
     &	\textbf{Index}	& \textbf{Name} &	 &	 (\textbf{sec}) \\ \hline
     
     & & & Sustain sound at \underline{vowel} &  \\
     & 01& beet& b\underline{ee}t&\multirow{13}{*}{7 (x13)}\\
     & 02& bit& b\underline{it}&\\
     & 03& bait& b\underline{ai}t&\\
     & 04& bet& b\underline{e}t&\\
     & 05& bat& b\underline{a}t&\\
     Vowels & 06& pot&p\underline{o}t &\\
     in /bVt/& 07& but& b\underline{u}t&\\
     & 08& bought& b\underline{oug}ht&\\
     & 09& boat& b\underline{oa}t&\\
     & 10& boot&b\underline{oo}t &\\
     & 11& put& p\underline{u}t&\\
     & 12& bird& b\underline{ir}d&\\
     & 13& abbot& abb\underline{o}t&\\ \hline
   
     & & & Sustain sound at \underline{consonant} & \multirow{15}{*}{7 (x14)} \\
    &14&afa&a\underline{f}a as in \underline{f}ood&\\
    &15&ava&a\underline{v}a as in \underline{v}oice&\\
    &16&atha\_thing&a\underline{th}a as in \underline{th}ing&\\
    &17&atha\_this&a\underline{th}a as in \underline{th}is&\\
    &18&asa&a\underline{s}a as in \underline{s}oap&\\
    Consonants &19&aza&a\underline{z}a as in \underline{z}ipper&\\
    in symmetric&20&asha&a\underline{sh}a as in \underline{sh}oe&\\
    /VCV/	 &21&agea&a\underline{ge}a as in bei\underline{ge}&\\
    &22&aha&a\underline{h}a as in \underline{h}appy&\\
    &23&ama&a\underline{m}a as in ri\underline{m}&\\
    &24&ana&a\underline{n}a as in pi\underline{n}&\\
    &25&anga&a\underline{ng}a as in ri\underline{ng}&\\
    &26&ala&a\underline{l}a as in \underline{l}ate&\\
    &27&ara&	a\underline{r}a as in \underline{r}ope&\\\hline

    &\multirow{2}{*}{28}&\multirow{2}{*}{breathe} 
    &  Breathe normally with your mouth & \multirow{10}{*}{7 (x6)} \\
    &&&closed, resting. &\\
    & 29 & clench & Clench your teeth and hold.&\\
    &\multirow{2}{*}{30}&\multirow{2}{*}{tongue} 
    &  Stick your tongue out &\\
    &&& as far as you can, and hold. &\\
    Postures	&\multirow{3}{*}{31}&\multirow{3}{*}{yawn}
    &  Pull back your tongue as far into & \\
    &&&  the mouth as you can, & \\
    &&& and hold (like yawning).&\\
    &\multirow{2}{*}{32}&\multirow{2}{*}{tip}
    &  Raise the tip of your tongue to & \\
    &&&  the middle of your palate, and hold.&\\
    &33&hold&Hold your breath.&\\ \hline
    \multicolumn{4}{c}{Repetition of the above tasks} & 7 (x33) \\ \hline

    \end{tabular}
    \label{tab4}
    \end{adjustwidth}
\end{table}

\subsection*{RT-MRI Acquisition}
RT-MRI acquisition was performed using a 13-interleaf spiral-out spoiled gradient-echo pulse sequence \cite{28}. This is an efficient scheme for sampling MR measurements, each with the different initial angle being interleaved by the bit-reversed order in time \cite{29}. The 13 spiral interleaves, when collected together, fulfil the Nyquist sampling rate. Imaging parameters were: repetition time = 6.004 ms, echo time = 0.8 ms, field-of-view (FOV) = 200 $\times$ 200 mm$^2$, slice thickness = 6 mm, spatial resolution = 2.4 $\times$ 2.4 mm$^2$ (84 $\times$ 84 pixels), receiver bandwidth = $\pm$ 125 kHz, flip angle = 15$^{\circ}$. Imaging was performed in the mid-sagittal plane, which was prescribed using a real‐time interactive imaging platform \cite{30} (RT-Hawk, Heart Vista, Los Altos, CA). Real-time visualization was implemented within the custom platform by using a sliding window gridding reconstruction \cite{31} to ensure subject’s compliance with stimuli and to detect substantial head movement. 

Acquisition was divided into 20-40 second task intervals presented in Table \ref{tab2}, each followed by a pause of the same duration so as to allow enough a brief break for the subject prior to the next task and to avoid gradient heating. 

\subsection*{RT-MRI Reconstruction}
The dataset provides one specific image reconstruction that has been widely used in speech production research \cite{21,28}. This image reconstruction involves optimizing the following cost function:

\begin{equation} \label{eq1}
    {min}_m \|Am-d\|_2^2+ \lambda \|\bigtriangledown_t m\|_1
\end{equation} 
where A represents the encoding matrix that models for the non-uniform fast Fourier transform and the coil sensitivity encoding, m is the dynamic image time series to be reconstructed, d is the acquired multiple-coil raw data, $\lambda$ is the regularization parameter, $\bigtriangledown_t$ is the temporal finite difference operator, and $\|\cdot\|_2^2$ and $\|\cdot\|_1$ are l$_2$ and l$_1$ norms, respectively. The coil sensitivity maps were estimated using the Walsh method \cite{32} from temporally combined coil images. The regularization term encourages voxels to be piecewise constant along time. This regularization has been successfully applied to speech RT-MRI \cite{21,33,34} and a variety of other dynamic MRI applications where the primary features of interest are moving tissue boundaries \cite{35,36,37}.

Our reference implementation solves Equation (\ref{eq1}) using nonlinear conjugate gradient algorithm with Fletcher-Reeves updates and backtracking line search \cite{38}. The algorithm was terminated either at 150 maximum iterations or the step size fell below $<$ 1e-5 during line search. Reference images were reconstructed using 2 spiral arms per time frame, resulting in a temporal resolution of 83.28 frames per second. This temporal resolution is enough to capture important articulator motions1, but reconstruction at different temporal resolutions is also possible with the provided dataset by adjusting the number of spiral arms per time frame. The algorithm was implemented in both MATLAB (The MathWorks, Inc., Natick, MA) and Python (Python Software Foundation, \url{https://www.python.org/}). The provided image reconstruction was performed in MATLAB 2019b on a Xeon E5-2640 v4 2.4GHz CPU (Intel, Santa Clara, CA) and a Tesla P100 GPU (Nvidia, Santa Clara, CA). Reconstruction time was 160.69 $\pm$ 1.56 ms per frame. Parameter selection of $\lambda$ is described in the technical validation section below. 

\subsection*{Audio Data}
One main microphone was positioned $\sim$ 0.5 inch away from the subject’s mouth. The microphone signal was sampled at 100 kHz each. The data were recorded on a laptop computer using a National Instruments NI-DAQ 6036E PCMCIA card. The audio sample clock was hardware-synchronized to the MRI scanner’s 10 MHz master clock. The audio recording was started and stopped using the trigger pulse signal from the scanner. The real-time audio data acquisition routine was written in MATLAB. Audio was first low-pass filtered and decimated to a sampling frequency of 20 kHz. The recorded audio was then enhanced using a normalized least-mean-square noise cancellation method \cite{23} and was aligned with the reconstructed MRI video sequence to aid linguistic analysis. 

\subsection*{3D Volumetric MRI of Sustained Sounds}
An accelerated 3D gradient-echo sequence with Cartesian sparse sampling was implemented to provide static high-resolution images of the full vocal tract during sustained sounds or postures. Imaging parameters were: repetition time = 3.8 ms, spatial resolution = 1.25 $\times$ 1.25 $\times$ 1.25 mm3, FOV = 200 $\times$ 200 $\times$ 100 mm3 (respectively in the anterior-posterior, superior-inferior, and left-right directions), image matrix size = 160 $\times$ 160 $\times$ 80, and flip angle = 5$^{\circ}$. The central portion (40 $\times$ 20) of the k$_y$-k$_z$ space was fully sampled to estimate the coil sensitivities from the data itself. The outer portion of the k$_y$-k$_z$ space was sampled using a sparse Poisson-Disc sampling pattern, which together resulted in 7-fold net acceleration, and a total scan time of 7 seconds. 

Data were acquired while the subjects sustained for 7 seconds a sound from the full set of American English vowels and continuant consonants (Table 4). The stimuli were presented to the subject via a projector-mirror setup, and upon hearing a “GO” signal given by the scanner operator, a subject started to sustain the sound or posture; this was followed by the operator triggering the acquisition manually. Subjects sustained the postures as long as they could hear the scanner operating. A recovery time of 5-10 seconds was given to the subject between the stimuli.

Image reconstruction was performed off-line by a sparse-SENSE constrained reconstruction similar to the optimization problem shown in Equation (\ref{eq1}). In contrast to Equation (\ref{eq1}) , isotropic spatial total variation constraints were used \cite{28,39}. The reconstruction was achieved using the open-source Berkeley Advanced Reconstruction Toolbox (BART) \cite{40}; the image reconstruction was performed in MATLAB using GPU acceleration.

\subsection*{T2-Weighted MRI at Rest}
Fast spin echo-based sequence was performed to provide high-resolution T2-weighted images with fine detail anatomical characteristics of the vocal tract at rest position. Imaging was run to obtain full sweeps of the vocal tract in the axial, coronal, and sagittal orientations, for each of which the number of slices ranges from 29 to 70, depending on the size of the vocal tract. Imaging parameters were: repetition time = 4600 ms, echo time = 120 - 122 ms, slice thickness = 3 mm, in-plane FOV = 300 $\times$ 300 mm$^2$, in-plane spatial resolution = 0.5859 $\times$ 0.5859 mm$^2$, number of averages = 1, echo train length = 25, and scan time = approximately 3.5 minutes per orientation. 

\section*{Data Records}
This dataset is publicly available in \textit{figshare} \cite{41}. The total size of this dataset is approximately 966 GB. It contains (i) raw 2D sagittal-view RT-MRI data, reconstructed images and videos, and synchronized denoised audio, (ii) 3D volumetric MRI images, and (iii) T2-weighted MRI images. The data for subject XYZ is contained in folder with identifier subXYZ (e.g., {\fontfamily{cmss}\selectfont sub001/}) and organized into three main folders: 2D RT-MRI data are located in the subfolder {\fontfamily{cmss}\selectfont 2drt/} (e.g., {\fontfamily{cmss}\selectfont sub001/2drt/}), 3D volumetric images in {\fontfamily{cmss}\selectfont 3d/}, and T2-weighted images in {\fontfamily{cmss}\selectfont t2w/}. The contents and data structures of the dataset are detailed as follows.

\subsection*{RT-MRI}
Raw RT-MRI data are provided in the vendor-agnostic MRD format (previously known as ISMRMRD) \cite{42,43}, which stores k-space MRI measurements, k-space location tables, and sampling density compensation weights. Parameters for the acquisition sequence are contained in the file header information. In addition, this dataset includes reconstructed image data for each subject and task in HDF5 format, audio files in WAV format, and videos in MPEG-4 format. 

For each subject, RT-MRI raw data is contained in the subfolder {\fontfamily{cmss}\selectfont raw/} (e.g., {\fontfamily{cmss}\selectfont sub001/2drt/raw/}), reconstructed image data in {\fontfamily{cmss}\selectfont recon/}, co-recorded audio (after noise cancellation) in {\fontfamily{cmss}\selectfont audio/}, and reconstructed speech videos with aligned audio in {\fontfamily{cmss}\selectfont video/}. Table \ref{tab5} summarizes the data structure and naming conventions for this dataset. 

\renewcommand{\arraystretch}{1.3}
\begin{table}[!t]
    \begin{adjustwidth}{-0.37in}{-0.37in} 
    \caption{Naming of folders and files. Subject identifiers correspond to Table \ref{tab1}, column Subject ID. Stimulus indices correspond to Tables \ref{tab2} and \ref{tab4}, column Stimulus Index. Stimulus names correspond to Tables \ref{tab2} and \ref{tab4}, column Stimulus Name.}
    \centering
    \begin{tabular}{cccc}
    \hline
    \textbf{Category} &	\textbf{Data folder} &	\textbf{Filename convention} &	\textbf{Description} \\ \hline
    \multirow{12}{*}{\textbf{RT-MRI}} & {\fontfamily{cmss}\selectfont subXYZ/2drt/}	& {\fontfamily{cmss}\selectfont $<$subject-identifier$>$\_2drt\_$<$stimulus-index$>$} & Raw RT-MRI  \\
    & {\fontfamily{cmss}\selectfont raw/} & {\fontfamily{cmss}\selectfont \_$<$stimulus-name$>$\_r$<$repetition$>$\_raw.h5} & k-space data  \\ 
    & & (e.g., {\fontfamily{cmss}\selectfont sub001\_2drt\_01\_vcv1\_r1\_raw.h5})	& in MRD format \\ 
    \cline{2-4}
    
    &	{\fontfamily{cmss}\selectfont subXYZ/2drt/}	&{\fontfamily{cmss}\selectfont $<$subject-identifier$>$\_2drt\_$<$stimulus-index$>$} & 	Reconstructed  \\
    & {\fontfamily{cmss}\selectfont recon/} & {\fontfamily{cmss}\selectfont \_$<$stimulus-name$>$\_r$<$repetition$>$\_recon.h5} & RT-MRI image data \\
    &&&in HDF5 format \\
    \cline{2-4}
    &	{\fontfamily{cmss}\selectfont subXYZ/2drt/}	&{\fontfamily{cmss}\selectfont $<$subject-identifier$>$\_2drt\_$<$stimulus-index$>$} & Co-recorded\\
    & {\fontfamily{cmss}\selectfont audio/} & {\fontfamily{cmss}\selectfont \_$<$stimulus-name$>$\_r$<$repetition$>$\_audio.wav} & 	audio data  \\ 
    &&& in WAV format \\
    \cline{2-4}
    
    & {\fontfamily{cmss}\selectfont subXYZ/2drt/} & {\fontfamily{cmss}\selectfont $<$subject-identifier$>$\_2drt\_$<$stimulus-index$>$} &Videos of speech task \\ 
    & {\fontfamily{cmss}\selectfont video/} &  {\fontfamily{cmss}\selectfont \_$<$stimulus-name$>$\_r$<$repetition$>$\_video.mp4} & with aligned audio \\
    &&&in MPEG-4 format \\ \hline

    % 3D volumetric	
    & {\fontfamily{cmss}\selectfont subXYZ/3d/} & {\fontfamily{cmss}\selectfont $<$subject-identifier$>$\_3d\_$<$stimulus-index$>$} & Reconstructed \\ 
    & {\fontfamily{cmss}\selectfont recon/}	& {\fontfamily{cmss}\selectfont \_$<$stimulus-name$>$\_r$<$repetition$>$\_recon.mat} & volumetric image data \\ 
    \textbf{3D} & & (e.g., {\fontfamily{cmss}\selectfont sub001\_3d\_13\_abbot\_r1\_recon.mat})	  & in MAT format \\ 
    \cline{2-4}
    \textbf{volumetric}& {\fontfamily{cmss}\selectfont subXYZ/3d/} & {\fontfamily{cmss}\selectfont $<$subject-identifier$>$\_3d\_$<$stimulus\_index$>$} & Mid-sagittal \\ 
    & {\fontfamily{cmss}\selectfont snapshot/}	& {\fontfamily{cmss}\selectfont \_$<$stimulus\_name$>$\_r$<$repetition$>$\_snapshot.png} & slice image \\ 
    & &	  & in PNG format \\ \hline

    \textbf{T2-} & 	{\fontfamily{cmss}\selectfont subXYZ/t2w/ } &	{\fontfamily{cmss}\selectfont $<$orientation-index$><$slice-index$>$.dcm} & T2-weighted image \\
    \textbf{weighted} & {\fontfamily{cmss}\selectfont dicom/ }& (e.g., {\fontfamily{cmss}\selectfont 00010001.dcm})	&  in DICOM format \\\hline
    \end{tabular}
    \label{tab5}
    \end{adjustwidth}
\end{table}

\subsection*{3D Volumetric MRI of Sustained Sounds}
3D volumetric MRI data contains reconstructed image data and imaging parameters in MAT format (MATLAB, The MathWorks, Inc., Natick, MA) in the subfolder {\fontfamily{cmss}\selectfont recon/} (e.g., {\fontfamily{cmss}\selectfont sub001/3d/recon/}) and a mid-sagittal slice image in PNG format in the subfolder {\fontfamily{cmss}\selectfont snapshot/}. 

\subsection*{T2-Weighted MRI at Rest}
T2-weighted image data from axial, coronal, and sagittal orientations are stored in Digital Imaging and Communications in Medicine (DICOM) format in the subfolder {\fontfamily{cmss}\selectfont dicom/} (e.g., {\fontfamily{cmss}\selectfont sub001/t2w/dicom/}). 

The imaging DICOM files were de-identified using the Clinical Trial Processor (CTP) developed by the Radiological Society of North America (RSNA) \cite{44}. Specifically, data anonymization was completed using a command line tool developed in the Java programming language \cite{45}. All images followed the standardized DICOM format and some of the attributes were removed or modified to preserve privacy of the subjects, specifically: PatientName was modified to follow the subXYZ pattern, and the original study dates were shifted by the same offset for all subjects. 

\subsection*{Metadata}
We provide presentation slides that contain the experimental stimuli including scripts and pictures used for the visualization to the subjects, in PPT format in {\fontfamily{cmss}\selectfont Stimuli.ppt}. Demographic information for each subject is contained in XLSX format in {\fontfamily{cmss}\selectfont Subjects.xlsx}. This meta file contains sex, race, age (at the time of scan), height (cm), weight (kg), origin, birthplace, cities raised and lived, L1 (first language), L2 (second language, if any), L3 (third language, if any), and the first language and birthplace of each subject’s parents. 

Additionally, meta information for each subject and RT-MRI task is contained in JSON format in {\fontfamily{cmss}\selectfont metafile\textunderscore public\textunderscore$<$timestamp$>$.json}. For each subject, we include the following demographic information: 1) L1 (first language), 2) L2 (second language, if any), 3) sex (M/F), 4) age (years at the time of scan). Further, visual and audio quality assessment scores are provided using a 5-level Likert scale for each subject stratified into categories: off-resonance blurring (1, very severe; 2, severe; 3, moderate; 4, mild; 5, none), video SNR (1, poor; 2, fair; 3, good; 4, very good; 5, excellent), aliasing artifacts (1, very severe; 2, severe; 3, moderate; 4, mild; 5, none), and audio SNR (1, poor; 2, fair; 3, good; 4, very good; 5, excellent). Specifically, an MRI expert with 6 years of experience in reconstructing and reading speech RT-MRI examined all the RT-MRI videos reconstructed for each subject and assessed and scored their visual and audio quality. For each task of the individual subjects, the information about task’s index, name, existence of file, and notes taken during data inspection by inspectors are also contained in the meta file.

\section*{Technical Validation}
\subsection*{Data Inspection}
Four inspectors manually performed qualitative data inspection for the datasets. After reconstruction, all images were converted from HDF5 format to MPEG-4 format, at which time co-recorded audio (WAV format) was integrated. All qualitative data inspection was performed manually based on MPEG-4 format videos. Included in the dataset were files that met all of the following criteria:

\begin{enumerate}
    \item The video (MPEG-4) exists (no data handling failure). 
    \item The audio recording (WAV) exists (no data handling failure).
    \item The video and audio are synchronized (based on inspection by a human).
\end{enumerate} 

After the inspection, 75 subjects were included in this dataset. It should be noted that although three subjects (sub18, sub74, and sub75) present severe radio-frequency-interference artifacts that were later determined to be from a leak in the MRI radio-frequency cage, we included those three subjects in this dataset in the hopes of potentially facilitating development of a digital radio-frequency interference correction method. Files that did not exist (criteria 1) were annotated for each task and subject in the meta file \\ (i.e., {\fontfamily{cmss}\selectfont metafile\textunderscore public\textunderscore$<$timestamp$>$.json}). 

\begin{figure}[!hp]
    \begin{adjustwidth}{-0.1in}{-0.1in}
	\centering
	\includegraphics[width=1\textwidth]{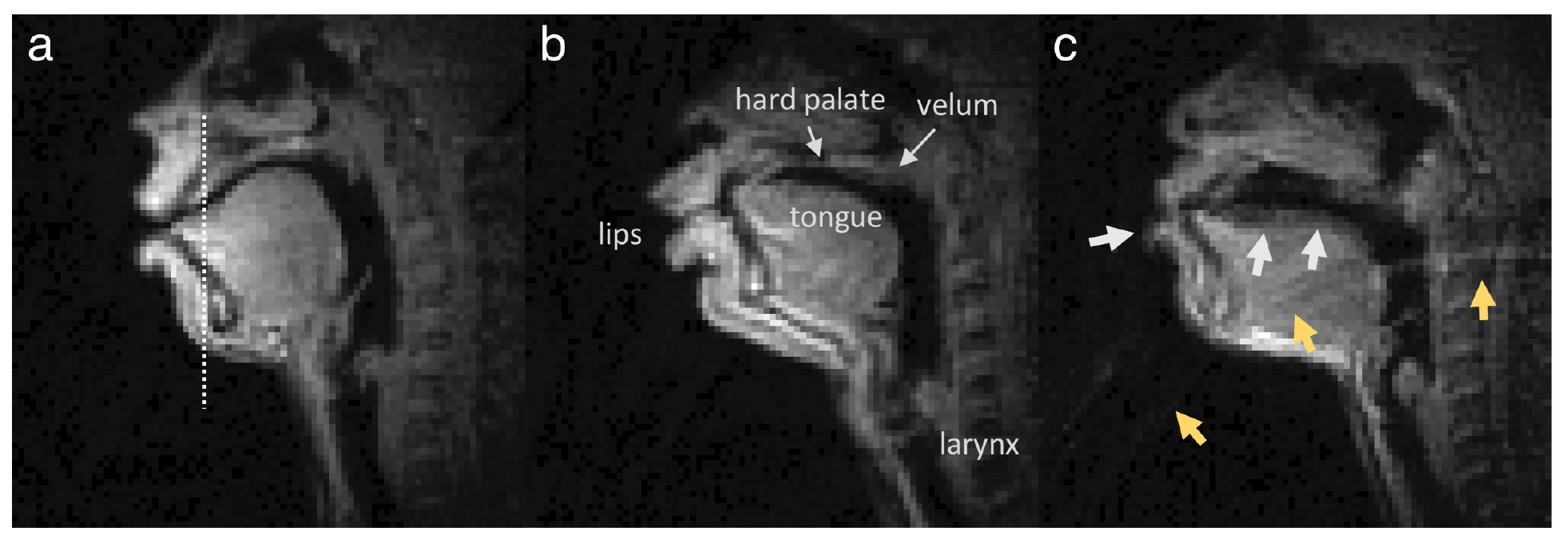}
	\caption{Typical data quality of 2D real-time speech imaging, shown in mid-sagittal image frames from three example subjects: (a) sub35 (male, 21yrs, native American English speaker), (b) sub51 (male, 33yrs, non-native speaker), (c) sub58 (female, 32yrs, non-native speaker). The mid-sagittal image frames depict the event of articulating the fricative consonant [$\theta$] in the word “uthu” (stimulus “vcv2”), where the tongue tip contacts the upper teeth. (a) and (b) are considered to have very high quality, based on high SNR and no noticeable artifact. (c) is considered to have moderate quality, based on good SNR and mild image artifacts; the white arrows point to blurring artifacts due to off-resonance while the yellow arrows point to ringing artifacts due to aliasing.}
	\label{fig2}
	\end{adjustwidth}
\end{figure}

Figure \ref{fig2} shows representative examples of the data quality from three subjects that are included in this dataset (sub35, sub41, and sub58). Note that all 75 subjects exhibited acceptable visualization of all soft tissue articulators. However, two types of artifacts were commonly observed:
\begin{enumerate}
    \item Blurring artifact due to off-resonance (green arrows, Figure \ref{fig2}c): This artifact appears as blurring or signal loss predominantly adjacent to air-tissue boundaries that surround soft tissue articulators. It is induced in spiral imaging by rapid changes of local magnetic susceptibility between the air and tissue. This artifact correction is still an active research area, including methods for a simple zeroth order frequency demodulation to advanced model-based \cite{14,46,47} and data-driven approaches \cite{15,48}. At present, we perform zeroth order correction during data acquisition as part of our routine protocol.
    
    \item Ringing artifact due to aliasing (yellow arrows, Figure \ref{fig2}c): This artifact appears as an arc-like pattern centered at the bottom-right corner outside the FOV. It is caused by a combination of gradient non-linearity and non-ideal readout low-pass filter when a strong signal source appears outside the unaliasing FOV. This artifact does not overlap with the articulator surfaces that are most important in the study of speech production. However, avoiding and/or correcting this artifact would improve overall image quality and potentially enable the use of a smaller FOV. 
\end{enumerate} 

\begin{figure}[!h]
    \begin{adjustwidth}{-0.1in}{-0.1in}
	\centering
	\includegraphics[width=1\textwidth]{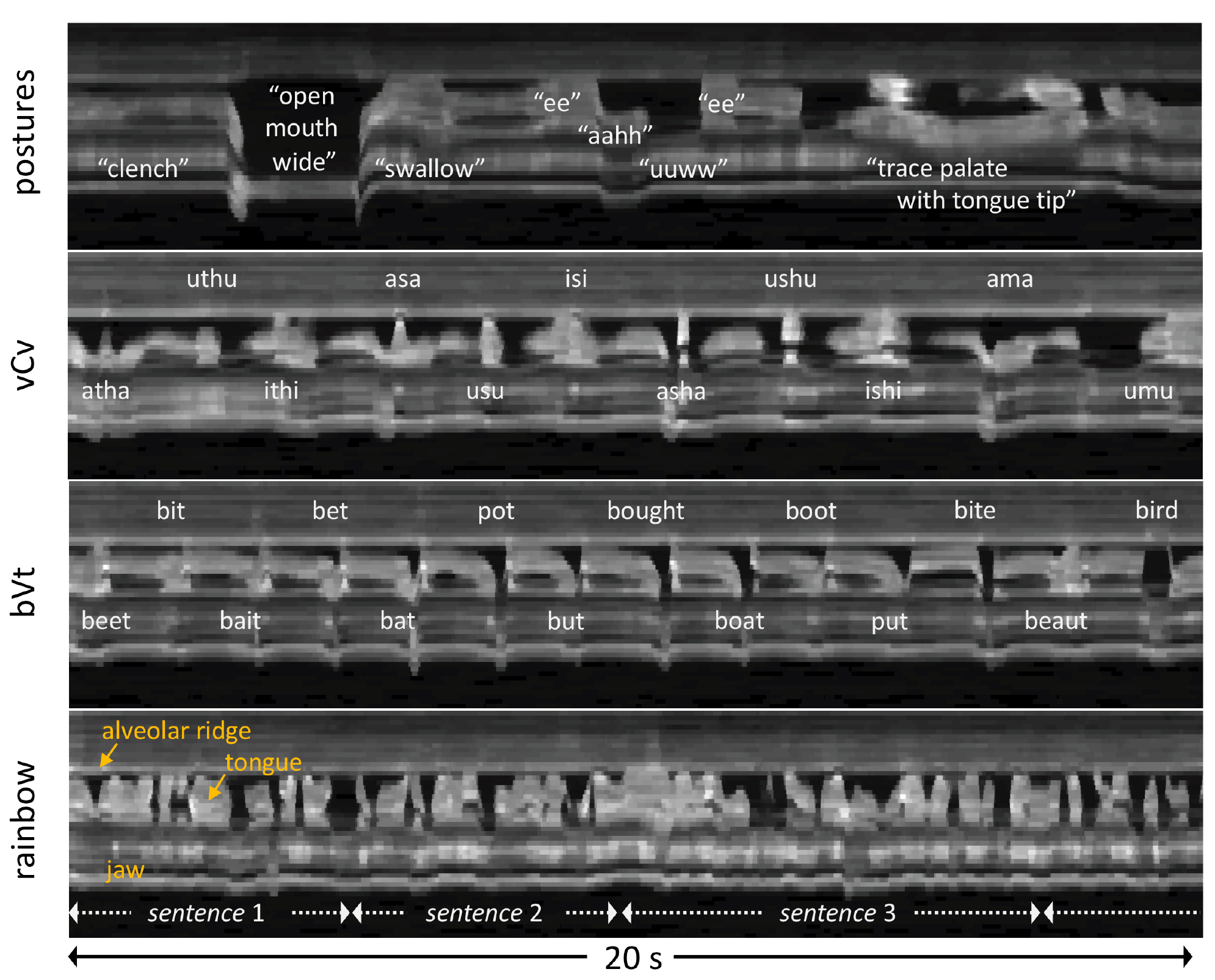}
	\caption{Image vs. time profiles during the first 20 seconds of four different stimuli for sub35. Profiles show the time evolution of the cut depicted by the dotted line in the image frame shown in Figure \ref{fig2}a (sub35). The rows visualize different examples of stimuli: “postures,” “vcv2,” “bvt,” and “rainbow” passage. The set of stimuli covers a wide range of articulator postures and tongue velocities.}
	\label{fig3}
	\end{adjustwidth}
\end{figure}

\begin{figure}[!h]
    \begin{adjustwidth}{-0.1in}{-0.1in}
	\centering
	\includegraphics[width=0.7\textwidth]{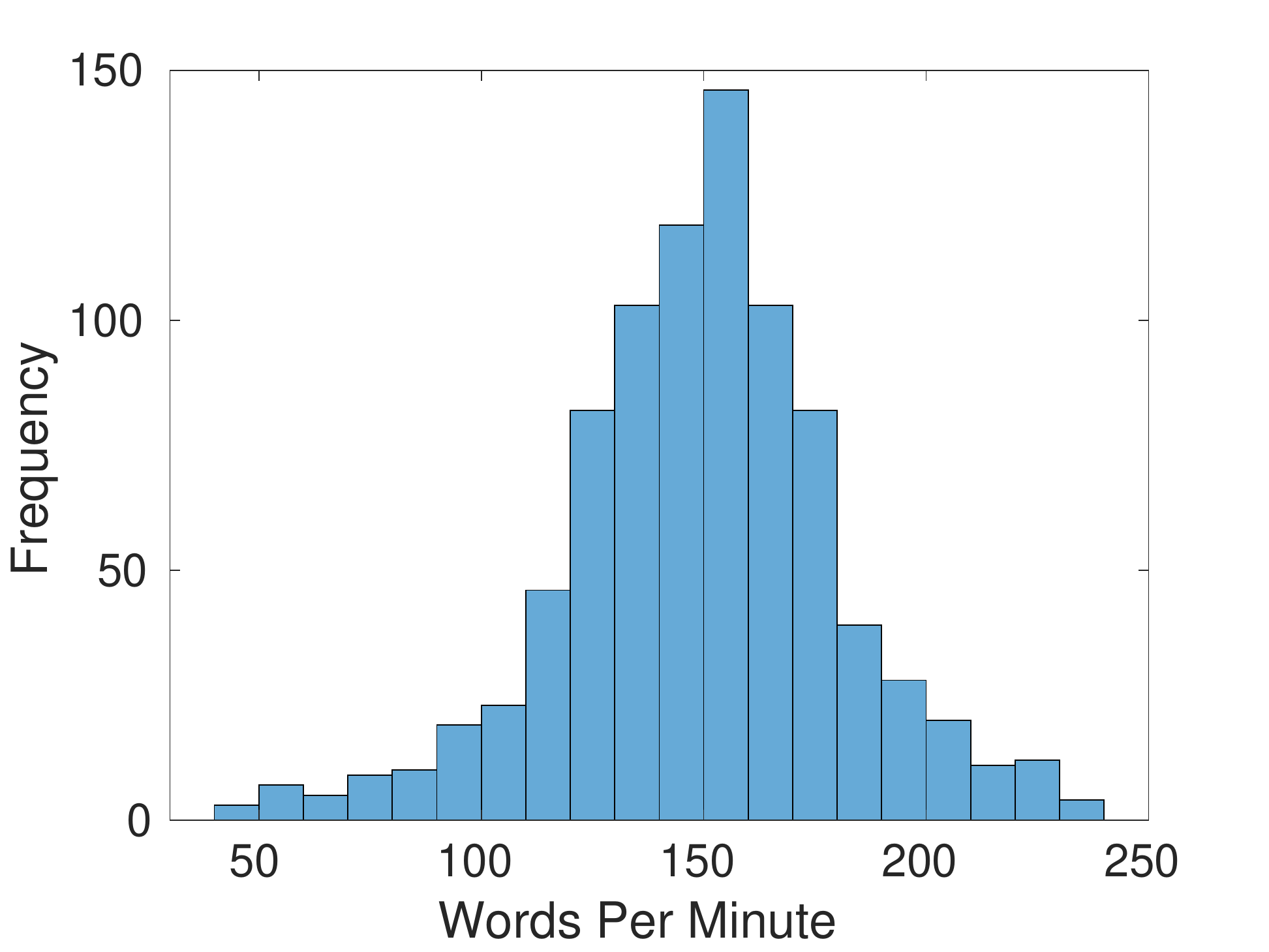}
	\caption{Histogram of words per minute during scripted speech stimuli including “shibboleth,” “rainbow,” “grandfather[1-2],” and “northwind[1-2].”}
	\label{fig4}
	\end{adjustwidth}
\end{figure}

Figure \ref{fig3} shows representative examples of the diverse speech stimuli that are included in this dataset. The intensity vs. time profiles visualize the first 20 seconds of four representative stimuli from sub35, shown in Figure \ref{fig2}a. These examples show the variety of patterns and speed of movement of the soft tissue articulators observed within a speaker as a function of the speech stimuli performed. Figure \ref{fig4} contains a histogram of the speaking rate in read sentences \cite{49}. The overall mean speaking rate was 149.2 ± 31.2 words per minute. The statistics are calculated from the read “shibboleth,” “rainbow,” “grandfather[1-2],” and “northwind[1-2]” stimuli for all subjects. These selected stimuli are composed of read full sentences. 

\begin{figure}[!h]
    \begin{adjustwidth}{-0.1in}{-0.1in}
	\centering
	\includegraphics[width=1\textwidth]{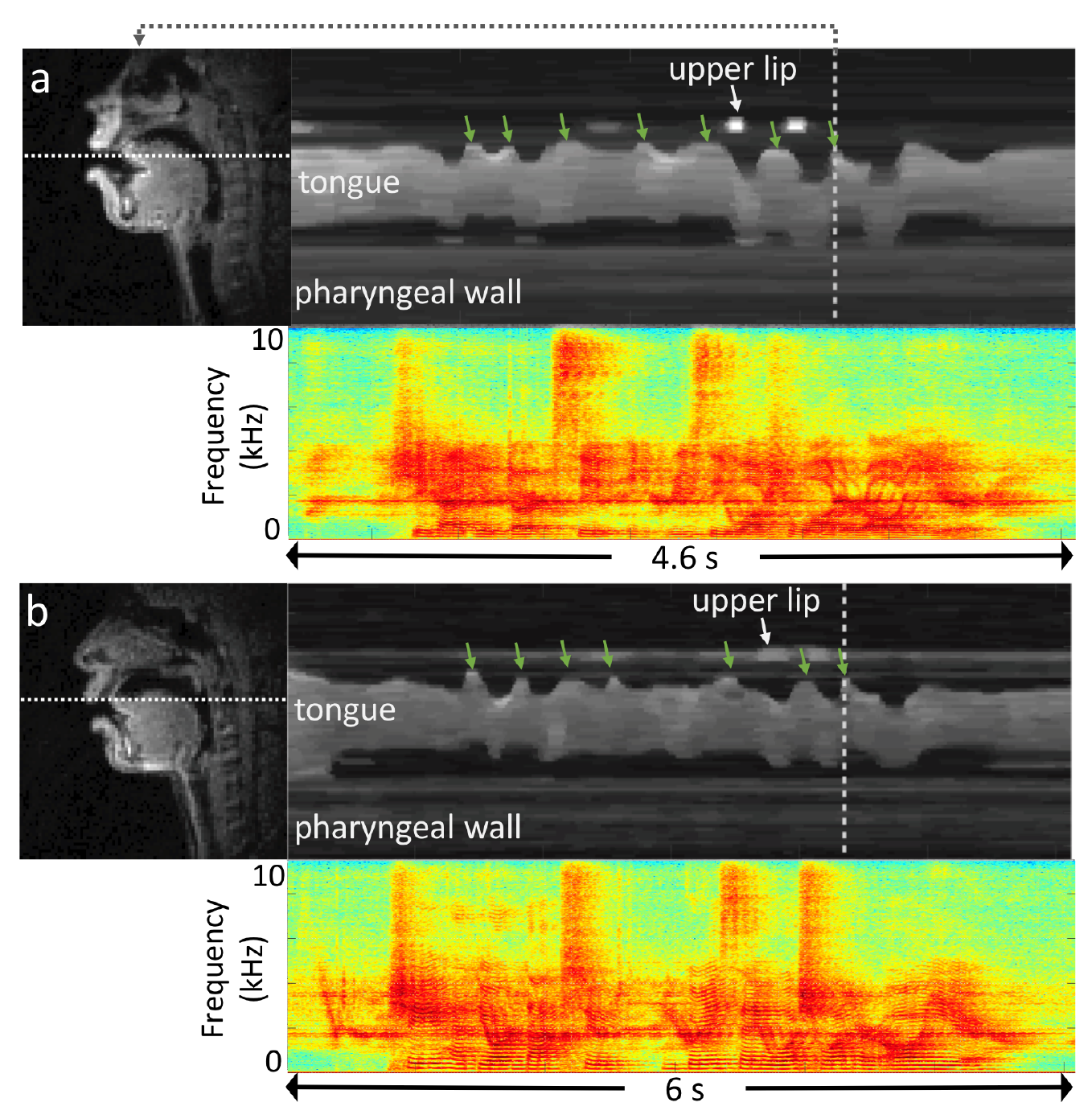}
	\caption{Variability in the articulation of the same sentence between two speakers: (a) sub35, (b) sub41. The time profile and audio spectrum correspond to the first sentence “She had your dark suit in greasy wash water all year” in the stimulus of “shibboleth” from each subject. The green arrows point to several noticeable time points at which the tongue tip contacts the upper teeth/alveolar ridge.}
	\label{fig5}
	\end{adjustwidth}
\end{figure}

Figure \ref{fig5} illustrates variability within the same speech stimuli across different speakers. The image time profiles correspond to the period of producing the first sentence “She had your dark suit in greasy wash water all year” in the stimulus “shibboleth” from sub35 and sub41. Although both speakers share the critical articulatory events (see green arrows), the timing and pattern vary depending on the subject. 

\subsection*{Regularization Parameter Selection for Image Reconstruction for RT-MRI}
\begin{figure}[!h]
    \begin{adjustwidth}{-0.1in}{-0.1in}
	\centering
	\includegraphics[width=1\textwidth]{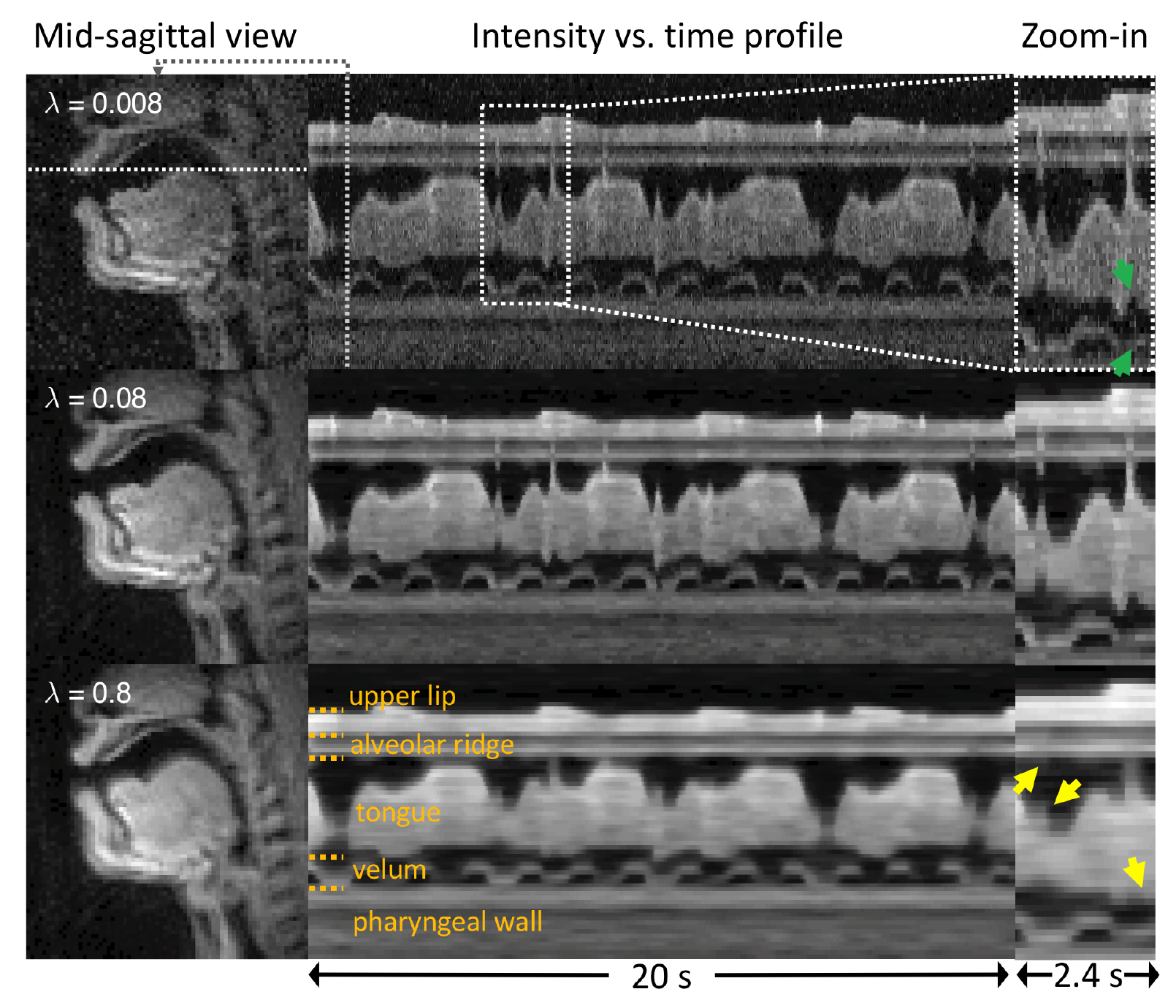}
	\caption{Illustration of the impact of reconstruction parameter $\lambda$ on image quality. Data are from sub15 (male, 26yrs, native American English speaker). (Left) Mid-sagittal image frames during speaking. (Middle) The intensity vs. time profiles for stimulus “vcv1.” (Right) Zoomed-in time profiles. Different rows correspond to different $\lambda$ values. For a smaller $\lambda$ (= 0.008C), the reconstruction shows a higher noise level and obscured articulatory event (green arrows), whereas for a higher $\lambda$ (= 0.8C), the noise level decreases but the temporal smoothing artifact is evident in regions indicated by yellow arrows. $\lambda$ = 0.08C yields an acceptable noise level while showing adequate temporal fidelity and therefore was selected for the optimal value for the reconstruction.}
	\label{fig6}
	\end{adjustwidth}
\end{figure}

We performed a parameter sweep and qualitative evaluation on a subset of the data to select a regularization parameter for the provided reconstructions. Ten stimuli from ten different subjects were randomly selected. The regularization parameter $\lambda$ was swept in the range between 0.008C and 1C in a logarithmic scale. Here, C represents the maximum intensity of the zero-filled reconstruction of the acquired data. Figure \ref{fig6} shows a representative example of the impact of $\lambda$ on the reconstructed image quality. A small $\lambda$ (= 0.008C) exhibits high noise level in the reconstruction (top, Figure \ref{fig6}). A higher $\lambda$ (= 0.8C) reduces the noise level but exhibits unrealistic temporal smoothing as shown in the intensity vs. time profiles (yellow arrows, Figure \ref{fig6}). The optimal parameter 0.08C was selected by consensus among four experts in the area of MRI image reconstruction and/or speech production RT-MRI. Once the $\lambda$ was optimized for the subset of the data, reconstruction was performed for all datasets. We have empirically observed that the choice of $\lambda$ appears robust across all of the datasets.

\section*{Usage Notes}
Several papers have been published by our group in which methods are directly applied to a subset of this dataset for the reconstruction and artifact correction of the RT-MRI data. These include auto-calibrated off-resonance correction \cite{14}, deblurring using convolutional neural networks \cite{15}, aliasing artifact mitigation \cite{50} (the artifact marked by the yellow arrows in Figure \ref{fig2}c), and super-resolution reconstruction [under review]. 

Additionally, several tools have been developed at our site for the analysis and modelling of reconstructed real-time MRI data. These include a graphical user interface for efficient visual inspection \cite{17} and implementations of grid-based tracking of air-tissue boundaries \cite{51,52,53}, region segmentation and factor analysis \cite{54,55,56}, neural network-based edge detection \cite{57,58}; region of interest (ROI) analysis \cite{59,60,61} and centroid tracking \cite{62}. Some of these tools are also made available; see code repository at \url{https://github.com/usc-mrel/usc\_speech\_mri.git}, as well as Ramanarayanan et al \cite{3} and Toutios et al \cite{63} for detailed reviews.

\section*{Code Availability}
This dataset is accompanied by a code repository (\url{https://github.com/usc-mrel/usc\_speech\_mri.git}) that contains examples of software and parameter configurations necessary to load and reconstruct the raw RT-MRI in MRD format. Specifically, the repository contains demonstrations to illustrate and replicate results of Figures 2-6. Code samples are available in MATLAB and Python programming languages. All software is provided free to use and modify under the MIT license agreement.

\section*{Acknowledgements}
This work was supported by NSF Grant 1514544, NSF Grant 1908865, and NIH Grant R01DC007124.  

\section*{Author contributions}

\textbf{Y.Lim}: wrote the manuscript and collected and curated data\\
\textbf{A.T.}: led the data collection and curation\\
\textbf{Y.B.}: contributed to data curation and the manuscript writing\\
\textbf{Y.T.}: contributed to data curation and the manuscript writing\\
\textbf{S.G.L.}: developed data acquisition protocols and collected data\\
\textbf{C.V.}: developed data acquisition protocols and collected data\\
\textbf{T.S.}: collected and curated data\\
\textbf{M.O.}: collected and curated data \\
\textbf{S.H.}: collected and curated data \\
\textbf{Y.Lee}: collected data \\
\textbf{W.C.}: collected data \\
\textbf{J.T.}: collected data \\
\textbf{M.L.M.}: collected data \\
\textbf{C.S.}: collected data\\
\textbf{B.G.}: curated data\\
\textbf{L.G.}: designed experimental stimuli \\
\textbf{D.B.}: designed experimental stimuli\\
\textbf{K.N.}: managed the project; developed data acquisition protocols; contributed to data curation\\
\textbf{S.N.}: conceived the project; managed the project; contributed to data curation

All authors contributed to the paper preparation, reviewed drafts of the paper, and approved of the final version.

\section*{Competing interests}
The authors declare no competing interests.

\clearpage

\section*{Figures and figures legends}

\def\figurename{Figure }

\def\figurename{Supplemental Figure }
\renewcommand{\thefigure}{1}
\begin{figure}[!h]
    \begin{adjustwidth}{-0.1in}{-0.1in}
	\centering
	\includegraphics[width=1\textwidth]{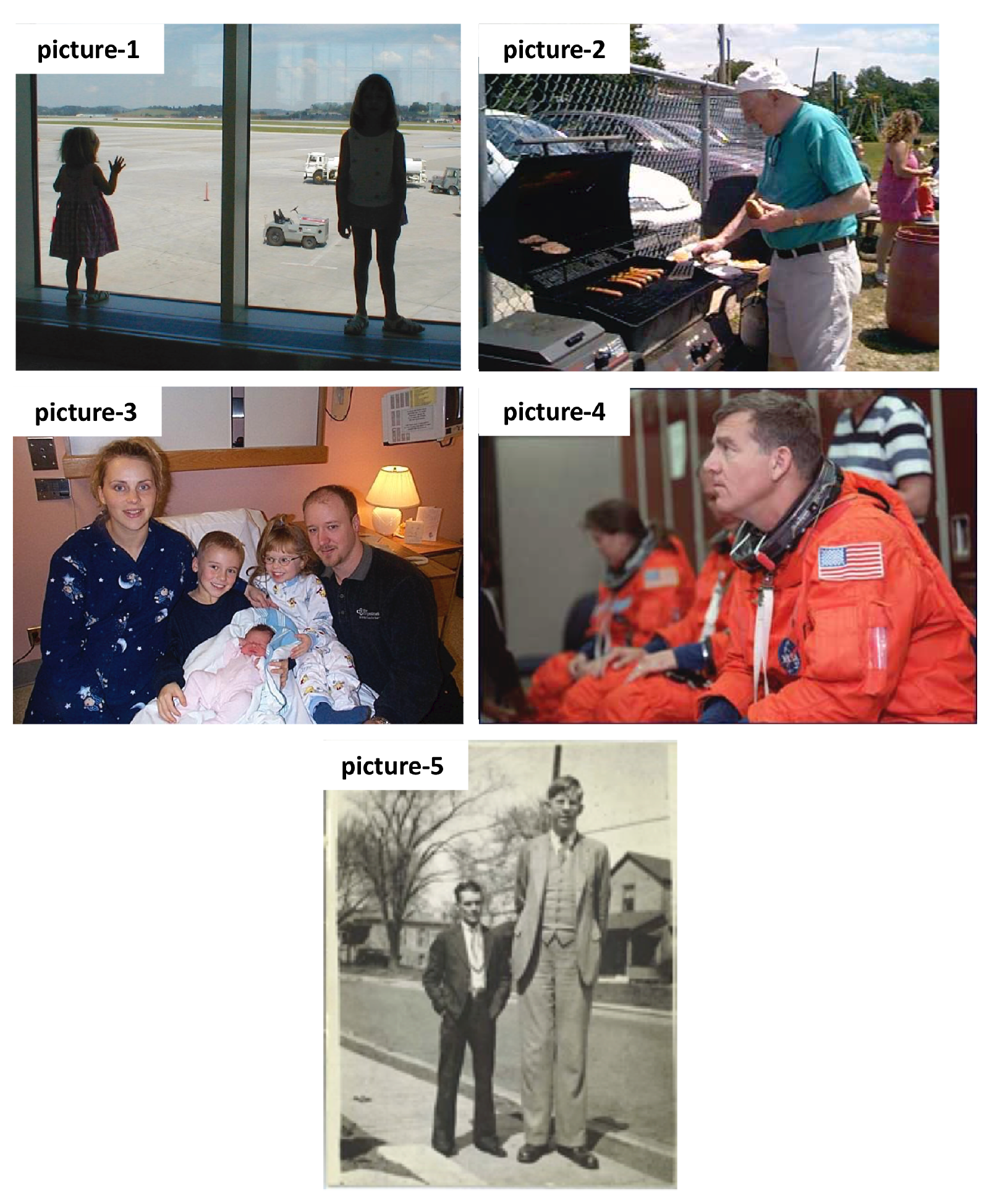}
	\caption{Photographs corresponding to speech experiment stimuli of picture1 to picture5. Photo source: \\
	\url{https://writefix.com/?page_id=411} (picture1), \\
	\url{https://writefix.com/?page_id=438} (picture2), \\
	\url{https://writefix.com/?page_id=443} (picture3), \\
	\url{https://writefix.com/?page_id=400} (picture4), \\
	\url{https://farmvilleherald.com/2020/03/the-worlds-biggest-man-visits-farmville/} (picture5)}
	\label{supfig1}
	\end{adjustwidth}
\end{figure}


\clearpage
\begin{thebibliography}{63}
\bibitem{1} Lingala, S. G., Sutton, B. P., Miquel, M. E. $\&$ Nayak, K. S. Recommendations for real-time speech MRI. \textit{J. Magn. Reson. Imaging} \textbf{43}, 28--44 (2016).

\bibitem{2}
Scott, A. D., Wylezinska, M., Birch, M. J. $\&$ Miquel, M. E. Speech MRI: Morphology and function. \textit{Phys. Medica} \textbf{30}, 604--618 (2014).

\bibitem{3}
Ramanarayanan, V. et al. Analysis of speech production real-time MRI. \textit{Comput. Speech. Lang.} \textbf{52}, 1--22 (2018).

\bibitem{4}
Hagedorn, C. et al. Engineering Innovation in Speech Science: Data and Technologies. \textit{Perspect. ASHA Spec. Interes. Groups} 4, 411--420 (2019).

\bibitem{5}
Bresch, E., Kim, Y. C., Nayak, K., Byrd, D. $\&$ Narayanan, S. Seeing speech: Capturing vocal tract shaping using real-time magnetic resonance imaging. \textit{IEEE Signal Process. Mag.} \textbf{25}, 123--129 (2008).
\bibitem{6}
Nayak, K. S., Lim, Y., Campbell-Washburn, A. E. $\&$ Steeden, J. Real-Time Magnetic Resonance Imaging. \textit{J. Magn. Reson. Imaging}, (2020). \doi{10.1002/jmri.27411}

\bibitem{7}
Marcus, D. S., Wang, T. H., Parker, J., Csernansky $\&$ J. G. Open access series of imaging studies (OASIS): Cross-sectional MRI data in young, middle aged, nondemented, and demented older adults. \textit{J. Cogn. Neurosci.} \textbf{19}, 1498--1507 (2007).

\bibitem{8}
Souza, R. et al. An open, multi-vendor, multi-field-strength brain MR dataset and analysis of publicly available skull stripping methods agreement. \textit{Neuroimage} \textbf{170}, 482--494 (2018).

\bibitem{9}
Knoll, F. et al. fastMRI: A Publicly Available Raw k-Space and DICOM Dataset of Knee Images for Accelerated MR Image Reconstruction Using Machine Learning. \textit{Radiol. Artif. Intell.} \textbf{2}, e190007 (2020).

\bibitem{10}
Chen, C. et al. OCMR (v1.0)--Open-access multi-coil k-space dataset for cardiovascular magnetic resonance imaging. arXiv:2008.03410 (2020).

\bibitem{11}
\url{http://mridata.org/}

\bibitem{12}
Knoll, F. et al. Advancing machine learning for MR image reconstruction with an open competition: Overview of the 2019 fastMRI challenge. \textit{Magn. Reson. Med.} \textbf{84}, 3054--3070 (2020).

\bibitem{13}
Sutton, B. P., Conway, C. A., Bae, Y., Seethamraju, R. $\&$ Kuehn, D. P. Faster dynamic imaging of speech with field inhomogeneity corrected spiral fast low angle shot (FLASH) at 3 T. \textit{J. Magn. Reson. Imaging} \textbf{32}, 1228--1237 (2010).

\bibitem{14}
Lim, Y., Lingala, S. G., Narayanan, S. S. $\&$ Nayak, K. S. Dynamic off-resonance correction for spiral real-time MRI of speech. \textit{Magn. Reson. Med.} \textbf{81}, 234--246 (2019).

\bibitem{15}
Lim, Y., Bliesener, Y., Narayanan, S. S. $\&$ Nayak, K. S. Deblurring for spiral real-time MRI using convolutional neural network. \textit{Magn. Reson. Med.} \textbf{84}, 3438--3452 (2020). 

\bibitem{16}
Töger, J. et al. Test–retest repeatability of human speech biomarkers from static and real-time dynamic magnetic resonance imaging. \textit{J. Acoust. Soc. Am.} \textbf{141}, 3323–3336 (2017).

\bibitem{17}
Narayanan, S. et al. Real-time magnetic resonance imaging and electromagnetic articulography database for speech production research (TC). \textit{J. Acoust. Soc. Am.} \textbf{136}, 1307--1311 (2014).

\bibitem{18}
Kim, J. et al. USC-EMO-MRI corpus: An emotional speech production database recorded by real-time magnetic resonance imaging. In \textit{Proc. the 10th Int. Semin. Speech Prod.}, 5--8 (2014).

\bibitem{19}
Sorensen, T. et al. Database of volumetric and real-time vocal tract MRI for speech science. In \textit{Proc. Annu. Conf. Int. Speech Commun. Assoc. (INTERSPEECH)}, 645--649 (2017). 

\bibitem{20}
Toutios, A. $\&$ Narayanan, S. S. Advances in real-time magnetic resonance imaging of the vocal tract for speech science and technology research. \textit{APSIPA Trans. Signal Inf. Process.} \textbf{5}, e6 (2016).

\bibitem{21}
Lingala, S. G. et al. A fast and flexible MRI system for the study of dynamic vocal tract shaping. \textit{Magn. Reson. Med.} \textbf{77}, 112--125 (2017).

\bibitem{22}
Lingala, S. G. et al. State-of-the-art MRI protocol for comprehensive assessment of vocal tract structure and function. In \textit{Proc. Annu. Conf. Int. Speech Commun. Assoc. (INTERSPEECH)}, 475--479 (2016).

\bibitem{23}
Bresch, E., Nielsen, J., Nayak, K. $\&$ Narayanan, S. Synchronized and noise-robust audio recordings during realtime magnetic resonance imaging scans. \textit{J. Acoust. Soc. Am.} \textbf{120}, 1791--1794 (2006).

\bibitem{24}
Garofolo, J. S., Lamel, L. F., Fisher, W. M., Fiscus, J. G. $\&$ Pallett, D. S. DARPA TIMIT acoustic-phonetic continous speech corpus CD-ROM. NIST speech disc 1-1.1. \textit{NASA STI/Recon Tech. Rep. N}, 27403 (1993).

\bibitem{25}
Fairbanks, F. The Rainbow Passage. \textit{Voice and Articulation Drillbook} 2nd edn. (New York: Harper $\&$ Row., 1960) 124--139.

\bibitem{26}
Darley, F. L., Aronson, A. E. $\&$ Brown, J. R. Motor Speech Disorders. (Saunders, 1975).

\bibitem{27}
Smith, C. L. Handbook of the International Phonetic Association: A guide to the use of the International Phonetic Alphabet (Cambridge University Press, 1999).

\bibitem{28}
Lingala, S. G. et al. State-of-the-art MRI protocol for comprehensive assessment of vocal tract structure and function. In \textit{Proc. Annu. Conf. Int. Speech Commun. Assoc. (INTERSPEECH)}, 475--479 (2016).

\bibitem{29}
Kerr, A. B. et al. Real-time interactive MRI on a conventional scanner. \textit{Magn. Reson. Med.} \textbf{38}, 355--367 (1997).

\bibitem{30}
Santos, J. M., Wright, G. A. $\&$ Pauly, J. M. Flexible real-time magnetic resonance imaging framework. In \textit{Proc. Annu. Int. Conf. IEEE Eng. Med. Biol. Soc. (EMBS)}, 1048--1051 (2004).

\bibitem{31}
Narayanan, S. S., Nayak, K. S., Lee, S., Sethy, A. $\&$ Byrd, D. An approach to real-time magnetic resonance imaging for speech production. \textit{J. Acoust. Soc. Am.} \textbf{115}, 1771--1776 (2004).

\bibitem{32}
Walsh, D. O., Gmitro, A. F. $\&$ Marcellin, M. W. Adaptive reconstruction of phased array MR imagery. \textit{Magn. Reson. Med.} \textbf{43}, 682--690 (2000).

\bibitem{33}
Burdumy, M. et al. One-second MRI of a three-dimensional vocal tract to measure dynamic articulator modifications. \textit{J. Magn. Reson. Imaging} \textbf{46}, 94--101 (2017).

\bibitem{34}
Lim, Y. et al. 3D dynamic MRI of the vocal tract during natural speech. \textit{Magn. Reson. Med.} \textbf{81}, 1511--1520 (2019). 

\bibitem{35}
Bassett, E. C. et al. Evaluation of highly accelerated real-time cardiac cine MRI in tachycardia. \textit{NMR Biomed.} \textbf{27}, 175--182 (2014).

\bibitem{36}
Haji-Valizadeh, H. et al. Validation of highly accelerated real-time cardiac cine MRI with radial k-space sampling and compressed sensing in patients at 1.5T and 3T. \textit{Magn. Reson. Med.} \textbf{79}, 2745--2751 (2018).

\bibitem{37}
Steeden, J. A. et al. Real-time assessment of right and left ventricular volumes and function in children using high spatiotemporal resolution spiral bSSFP with compressed sensing. \textit{J. Cardiovasc. Magn. Reson.} \textbf{20}, 79 (2018).

\bibitem{38}
Lustig, M., Donoho, D. $\&$ Pauly, J. M. Sparse MRI: the application of compressed sensing for rapid MR imaging. \textit{Magn. Reson. Med.} \textbf{58}, 1182--1195 (2007).

\bibitem{39}
Kim, Y., Narayanan, S. $\&$ Nayak, K. Accelerated three-dimensional upper airway MRI using compressed sensing. \textit{Magn. Reson. Med.} \textbf{61}, 1434--1440 (2009).

\bibitem{40}
Uecker, M. et al. Berkeley Advanced Reconstruction Toolbox. In \textit{Proceedings of the Int. Soc. Magn. Reson. Med. (ISMRM)} \textbf{23}, 2486 (2015).

\bibitem{41}
Lim, Y. et al. A multispeaker dataset of raw and reconstructed speech production real-time MRI video and 3D volumetric images. figshare https://doi.org/10.6084/m9.figshare.13725546.v1 (2021).

\bibitem{42}
Inati, S. J. et al. ISMRM Raw data format: A proposed standard for MRI raw datasets. \textit{Magn. Reson. Med.} \textbf{77}, 411--421 (2017).

\bibitem{43}
https://ismrmrd.github.io/.

\bibitem{44}	
Radiological Society of North America I. CTP-The RSNA Clinical Trial Processor. Radiological Society of North America, Inc.

\bibitem{45}
\url{http://mircwiki.rsna.org/index.php?title=The\_DicomAnonymizerTool}.

\bibitem{46}
Fessler, J. A. et al. Toeplitz-Based Iterative Image Reconstruction for MRI With Correction for Magnetic Field Inhomogeneity. \textit{IEEE Trans. Signal. Process.} \textbf{53}, 3393--3402 (2005).

\bibitem{47}
Sutton, B. P., Conway, C. A., Bae, Y., Seethamraju, R. $\&$ Kuehn, D. P. Faster dynamic imaging of speech with field inhomogeneity corrected spiral fast low angle shot (FLASH) at 3 T. \textit{J. Magn. Reson. Imaging} \textbf{32}, 1228--1237 (2010).

\bibitem{48}	
Zeng, D. Y. et al. Deep residual network for off-resonance artifact correction with application to pediatric body MRA with 3D cones. \textit{Magn. Reson. Med.} \textbf{82}, 1398--1411 (2019).

\bibitem{49}
Jacewicz, E., Fox, R. A., O’Neill, C. $\&$ Salmons, J. Articulation rate across dialect, age, and gender. \textit{Lang. Var. Change} \textbf{21}, 233–256 (2009).

\bibitem{50}
Ye, T. et al. Aliasing artifact reduction in spiral real-time MRI. \textit{Magn. Reson. Med.} in press (2021). \doi{10.1002/mrm.28746}

\bibitem{51}
Proctor, M. I., Bone, D., Katsamanis, N. $\&$ Narayanan, S. Rapid Semi-automatic Segmentation of Real-time Magnetic Resonance Images for Parametric Vocal Tract Analysis. In \textit{Proc. Annu. Conf. Int. Speech Commun. Assoc. (INTERSPEECH)}, 1576--1579 (2010).

\bibitem{52}
Kim, J., Kumar, N., Lee, S. $\&$ Narayanan, S. Enhanced airway-tissue boundary segmentation for real-time magnetic resonance imaging data. In \textit{Proc. 10th Int. Semin. Speech Prod. (ISSP)}, 5--8 (2014).

\bibitem{53}
Kim, J., Toutios, A., Lee, S. $\&$ Narayanan, S. S. Vocal tract shaping of emotional speech. \textit{Comput. Speech Lang.} \textbf{64}, 101100 (2020). 

\bibitem{54}
Bresch, E. $\&$ Narayanan, S. Region segmentation in the frequency domain applied to upper airway real-time magnetic resonance images. \textit{IEEE Trans. Med. Imaging} \textbf{28}, 323--338 (2009).

\bibitem{55}
Toutios, A. $\&$ Narayanan, S. S. Factor analysis of vocal-tract outlines derived from real-time magnetic resonance imaging data. In \textit{Proc. 18th Int. Congress of Phonetic Sciences (ICPhS)} (2015)

\bibitem{56}
Sorensen, T., Toutios, A., Goldstein, L. $\&$ Narayanan, S. Task-dependence of articulator synergies. \textit{J. Acoust. Soc. Am.} \textbf{145}, 1504 (2019). 

\bibitem{57}
Somandepalli, K., Toutios, A. $\&$ Narayanan, S. S. Semantic edge detection for tracking vocal tract air-tissue boundaries in real-time magnetic resonance image. In \textit{Proc. Annu. Conf. Int. Speech Commun. Assoc. (INTERSPEECH)}, 631--635 (2017).

\bibitem{58}
Hebbar, S. A., Sharma, R., Somandepalli, K., Toutios, A. $\&$ Narayanan, S. Vocal Tract Articulatory Contour Detection in Real-Time Magnetic Resonance Images Using Spatio-Temporal Context. In \textit{IEEE International Conference on Acoustics, Speech and Signal Processing, (ICASSP)}, 7354--7358 (2020). 

\bibitem{59}
Lammert, A. C., Proctor, M. I. $\&$ Narayanan, S. S. Data-Driven Analysis of Realtime Vocal Tract MRI using Correlated Image Regions. In \textit{Proc. Annu. Conf. Int. Speech Commun. Assoc. (INTERSPEECH)}, 1572--1575 (2010).

\bibitem{60}
Lammert, A., Ramanarayanan, V., Proctor, M. $\&$ Narayanan, S. Vocal tract cross-distance estimation from real-time MRI using region-of-interest analysis. In \textit{Proc. Annu. Conf. Int. Speech Commun. Assoc. (INTERSPEECH)}, 959--962 (2013).

\bibitem{61}
Proctor, M. et al. Direct estimation of articulatory kinematics from real-time magnetic resonance image sequences. In \textit{Proc. Annu. Conf. Int. Speech Commun. Assoc. (INTERSPEECH)}, 281--284 (2011).

\bibitem{62}
Oh, M. $\&$ Lee, Y. ACT: An Automatic Centroid Tracking tool for analyzing vocal tract actions in real-time magnetic resonance imaging speech production data. \textit{J. Acoust. Soc. Am.} \textbf{144}, EL290--EL296 (2018).

\bibitem{63}
Toutios, A., Byrd, D., Goldstein, L. $\&$ Narayanan, S. Advances in vocal tract imaging and analysis. \textit{The Routledge Handbook of Phonetics} (Routledge, 2019).
 
 




% LNCS Homepage, \url{http://www.springer.com/lncs}. Last accessed 4
\end{thebibliography}
\end{document}